\documentstyle[aps,twocolumn,epsf,floats]{revtex}

\begin{document}

\draft

\date{April 2000}

\title{Coherent backscattering effect on wave dynamics in a random medium}
\author{H. Schomerus, K. J. H. van Bemmel, and C. W. J. Beenakker
}
\address{Instituut-Lorentz, Universiteit Leiden, P.O. Box 9506, 2300 RA
Leiden, The Netherlands}
\twocolumn[
\widetext
\begin{@twocolumnfalse}

\maketitle

\begin{abstract}
A dynamical effect of coherent backscattering is predicted theoretically
and supported by computer simulations: The distribution of single-mode
delay times of waves reflected by a disordered waveguide depends on
whether the incident and detected modes are the same or not. The change
amounts to a rescaling of the distribution by a factor close to
$\sqrt{2}$. This effect appears only if the length of the waveguide
exceeds the localization length; there is no effect of coherent
backscattering on the delay times in the diffusive regime.
\end{abstract}

\end{@twocolumnfalse}
]

\narrowtext

Coherent backscattering refers to the systematic constructive
interference of waves reflected from a medium with randomly located
scatterers. The constructive interference occurs in a narrow cone around
the angle of incidence, and is a fundamental consequence of time-reversal
symmetry \cite{berk}. The resulting peak in the angular dependence 
of the reflected intensity is a generic wave effect: It has been
observed using light waves \cite{lw} and acoustic waves \cite{aw}, 
for classical and quantum scatterers \cite{cqs}, in passive and active media
\cite{med}.

These studies mainly addressed static properties. Dynamic aspects of wave
propagation in random media are now entering the focus of attention
\cite{vantiggelen:1999a,vantiggelen:1999b,LambertFalko,BeenBem},
and the work on acoustic waves \cite{aw} has started to study the
connection with the coherent backscattering effect.
The key observable in the dynamic experiments \cite{vantiggelen:1999a}
is the derivative $\phi'={\rm d}\phi/{\rm d}\omega$ of the phase $\phi$
of the wave amplitude with respect to the frequency $\omega$. 
The quantity $\phi'$ has the dimension of a time and is interpreted as
a {\em delay time}. Van Tiggelen {\it et\,al.}\ \cite{vantiggelen:1999b}
have developed a statistical theory for the distribution of the delay
time $\phi'$ and the intensity $I$ in a waveguide geometry (where angles
of incidence are discretized as modes). Although the theory was worked
out mainly for the case of transmission, the implications for reflection
are that the distribution $P(\phi')$ does not depend on whether the
detected mode $n$ is the same as the incident mode $m$ or not. This is
in contrast with $P(I)$, which is rescaled by a factor of $1/2$ when $n$
becomes equal to $m$---so that the mean $\bar I$ becomes twice as
large. Hence it appears that no coherent backscattering effect exists
for $P(\phi')$.

What we will demonstrate here is that this is true only if  
wave localization may be disregarded. Previous studies 
\cite{vantiggelen:1999a,vantiggelen:1999b} dealt with the diffusive
regime of waveguide lengths $L$ below the localization length $\xi$.
Here we consider the localized regime
$L>\xi$ (assuming that also the absorption length $\xi_{\rm a}>\xi$).
The distribution of reflected intensity is insensitive to the presence
or absence of localization, being given in both regimes by Rayleigh's
law:
\begin{equation}
\label{ray}
P(I)=\left\{\begin{array}{cc} Ne^{-N I} & \mbox{ if }n\neq m\;,\\
\frac{1}{2}Ne^{-N I/2} & \mbox{ if }n=m
\end{array}
\right.
\end{equation}
(for unit incident intensity).
In contrast, we find that the delay-time distribution changes markedly
as one enters the localized regime, decaying more slowly for large
$|\phi'|$. Moreover, a coherent backscattering effect appears: For $L >
\xi$ the peak of $P(\phi')$ is higher for $n=m$ than for $n\neq m$ by a
factor which is close to $\sqrt{2}$. We present a complete analytical
theory, compare it with numerical simulations, and offer a qualitative 
argument for this unexpected dynamical effect of coherent backscattering.

Let us begin with a more precise formulation of the problem.
We consider a disordered medium (mean free path $l$) in a waveguide geometry
(length $L$, with $N\gg 1$ propagating modes at frequency $\omega$)
and study the
correlator $\rho$  of the reflected wave amplitudes at
two nearby frequencies $\omega\pm\frac{1}{2}\delta\omega$,
\begin{equation}
\rho=r_{nm}(\omega+\case{1}{2}\delta\omega)r_{nm}^*(\omega-\case{1}{2}\delta\omega)
\;.
\label{eq:corr}
\end{equation}
The indices $n$ and $m$ specify the detected and incident mode,
respectively.
(We assume single-mode excitation and detection.)
The  amplitudes
$r_{nm}$
form the $N\times N$ reflection matrix $r$.
In the localized regime (localization length $\xi\simeq N l$
smaller than both $L$
and the absorption length $\xi_{\rm a}$), the matrix $r$ is
approximately unitary because transmission is negligibly small.
We assume time-reversal symmetry (no magneto-optical effects),
so that $r$ is also symmetric.
Following Genack {\it et\,al.}\ \cite{vantiggelen:1999a,vantiggelen:1999b} we 
define the single-mode (or single-channel) delay time $\phi'$ as
\begin{equation}
\phi'=\lim_{\delta \omega\to 0}
\frac{{\rm Im}\,\rho}{\delta\omega \, I}
\;,
\end{equation}
where $I=|r_{nm}(\omega)|^2$ is the intensity of the reflected wave
in the detected mode for unit incident intensity.
We seek the joint distribution function $P(I,\phi')$ in an ensemble of different
realizations of disorder.

The single-mode delay time $\phi'$ is a linear combination of the
Wigner-Smith \cite{fs} delay times
$\tau_i$ ($i=1,2,\ldots,N$), which are the eigenvalues of the matrix 
\begin{equation}
-i r^\dagger\frac{{\rm d}r}{{\rm d}\omega}=
U^\dagger\,{\rm diag}\,(\tau_1,\ldots,\tau_N) U
\;.
\end{equation}
(The matrix of eigenvectors $U$ is unitary for a unitary reflection
matrix.)
For small $\delta\omega$ we can expand 
\begin{equation}
r(\omega\pm\case{1}{2}\delta \omega)=U^{\rm T} U\pm
\case{1}{2}i\,\delta\omega\,
U^{\rm T}\,{\rm diag}\,(\tau_1,\ldots,\tau_N)\, U
\;,
\end{equation}
hence the relations
\begin{equation}
\label{eq:corr2}
\phi'={\rm Re}\,\frac{A_1}{A_0}\;,\quad I=|A_0|^2\;,\quad A_k=
\sum_i\tau_i^k u_i v_i\;.
\end{equation}
We have abbreviated $u_i=U_{im}$, $v_i=U_{in}$.

The distribution of the Wigner-Smith delay times for this problem was
determined recently \cite{BB}. In terms of the rates $\mu_i=1/\tau_i$
it has the form of the Laguerre ensemble of random-matrix theory,
\begin{equation}
P(\{\mu_i\})\propto \prod_{i<j}|\mu_i-\mu_j|\prod_{k}\Theta(\mu_k)
e^{-\gamma (N +1)\mu_k}
\;,
\label{eq:laguerre}
\end{equation}
where $\Theta(x)=1$ for $x>0$ and $0$ for $x<0$.
The parameter $\gamma=\alpha l/c$ (with wave velocity $c$)
equals the scattering rate times a numerical coefficient ($\alpha=\pi^2/4$,
$8/3$ for two, three-dimensional scattering).
Eq.\ (\ref{eq:laguerre}) extends the single-mode ($N=1$) result 
of Refs.\ \cite{Jayannavar,Heinrichs,Comtet} to any $N$.
The matrix $U$ is uniformly distributed in the unitary group.
We consider first the typical case $n\neq m$ of different incident and
detected modes. (The special case $n=m$ is addressed later.)
For $n\neq m$ the vectors ${\bf u}$ and ${\bf v}$
become uncorrelated in the large-$N$ limit, and
their elements become independent
Gaussian random numbers with vanishing mean and variance
$\langle |u_i^2|\rangle
=\langle |v_i^2|\rangle
=N^{-1}$.

It is convenient to work momentarily with
the weighted delay time
$W=\phi' I$ and to recover $P(I,\phi')$ from $P(I,W)$ at the end.
The characteristic function
$
\chi(p,q)=\left\langle e^{-ip I-iq W}\right\rangle
$
is the Fourier transform of $P(I,W)$.
The average 
$\langle \cdots\rangle$
is over the vectors ${\bf u}$ and ${\bf v}$ and over the set
of eigenvalues $\{\tau_i\}$. The average over one of the vectors, say
${\bf v}$, is easily carried out, because it
is a Gaussian integration. The result is a determinant, 
\begin{eqnarray}
&&\chi(p,q)=\left\langle\det(1+i H/N)^{-1}\right\rangle
\label{eq:chi1}
\;,
\\
&&H=p{\bf u}^*{\bf u}^{\rm T}+\case{1}{2}q (\bar{\bf u}^*
{\bf u}^{\rm T}+{\bf u}^*\bar{\bf u}^{\rm T})\;.
\end{eqnarray}
The Hermitian matrix $H$ is a sum
of dyadic products of the vectors ${\bf u}$ and $\bar{\bf u}$,
with  $\bar{u}_i=u_i \tau_i$,
and hence has only two non-vanishing eigenvalues $\lambda_+$ and
$\lambda_-$. Some straightforward linear algebra gives
\begin{equation}
\lambda_\pm=\case{1}{2}\left(q B_1+p\pm\sqrt{2pq B_1+q^2B_2+p^2}\right)
\;,
\end{equation}
where we have defined the spectral moments
\begin{equation}
B_k=\sum_i|u_i|^2\tau_i^k
\;.
\label{eq:ab}
\end{equation}
The resulting determinant is $\det(1+ H/N)^{-1}=(1+\lambda_+/N)^{-1}
(1+\lambda_-/N)^{-1}$, hence
\begin{eqnarray}
\chi(p,q)
=\left\langle
\left[1+\frac{ip}{N}+\frac{iq}{N}B_1+\frac{q^2}{4N^2}(B_2-B_1^2)\right]^{-1}
\right\rangle \;.
\end{eqnarray}
An inverse Fourier transform, followed by a change of variables from
$I$, $W$ to $I$, $\phi'$, gives
\begin{eqnarray}
&&P(I,\phi')=
\Theta(I)(N^3 I/\pi)^{1/2}e^{-N I}
\nonumber
\\
&&{}\quad{}\times
\left\langle
(B_2-B_1^2)^{-1/2}
\exp\left(-NI\frac{(\phi'-B_1)^2}{B_2-B_1^2}\right)
\right\rangle
.
\label{eq:iphi}
\end{eqnarray}
The average is over the spectral moments
$B_1$ and $B_2$, which depend on the
$u_i$'s and $\tau_i$'s via Eq.\ (\ref{eq:ab}).

This result in the localized regime 
is to be compared with the  result of diffusion theory
\cite{vantiggelen:1999a,vantiggelen:1999b},
\begin{eqnarray}
&&P_{\rm diff}(I,\phi')=
\Theta(I)(N^3I/\pi)^{1/2}e^{-N I}
\nonumber
\\
&&{}\quad{}\times
(Q \bar{\phi'}^2)^{-1/2}
\exp\left(-NI\frac{(\phi'-\bar{\phi'})^2}{Q\bar{\phi'}^2}\right)
\;.
\label{eq:iphifixed}
\end{eqnarray}
The constants are given by
$Q\simeq L/l$ and $\bar{\phi'}\simeq L/c$ up to numerical
coefficients of order unity \cite{note1}.
Comparison of Eqs.\ (\ref{eq:iphi}) and (\ref{eq:iphifixed}) shows that
the two distributions would be identical if statistical fluctuations in the
spectral moments $B_1$, $B_2$ could be ignored. 
However, as we shall see shortly,
the distribution $P(B_1,B_2)$ is very broad, so that
fluctuations can {\em not} be ignored. The large fluctuations are a
consequence of the high density of anomalously large
Wigner-Smith delay times $\tau_i$ 
in the Laguerre
ensemble  (\ref{eq:laguerre}), and are related to the penetration of the 
wave deep into the localized regions. The large $\tau_i$'s are
eliminated in the diffusive regime $L\lesssim\xi$.
Then $B_1$ and $B_2$ can be replaced by their ensemble averages,
and the Gaussian theory \cite{vantiggelen:1999a,vantiggelen:1999b}
is recovered. (The same applies if
the absorption length $\xi_{\rm a}\lesssim \xi$.)

To determine how the statistical fluctuations in the spectral moments
alter $P(I,\phi')$, we need the joint distribution $P(B_1,B_2)$. 
This can be calculated by applying the random-matrix technique
of Refs.\ \cite{pab1,pab2} to the Laguerre ensemble. The result is
\begin{eqnarray}
&&P(B_1,B_2)=\Theta(B_1)\Theta(B_2)\exp\left(-\frac{N B_1^2}{B_2}\right)
\nonumber\\ 
&&{}\times\left[
\frac{B_1^2\gamma N^3}{B_2^4}(B_2+\gamma N^2 B_1)
\exp\left(-\frac{2\gamma N}{B_1}\right)
\right.
\nonumber\\
&&\left.{}
-\frac{\gamma^3 N^5}{4B_2^5}(2B_2^2-4B_1^2B_2N+B_1^4N^2)
{\rm Ei}\,\left(-\frac{2\gamma N}{B_1}\right)
\right]
\;,
\label{eq:pab1}
\end{eqnarray}
where ${\rm Ei}\,(x)$ is the 
exponential-integral function.
The most probable values are
$B_1\sim \gamma N$,
$B_2\sim\gamma^2 N^3$, while the mean values $\langle B_1\rangle$, $\langle
B_2\rangle$ diverge---demonstrating the presence of large fluctuations.
The distribution $P(I,\phi')$ follows from 
Eq.\ (\ref{eq:iphi})
by integrating over $B_1$
and $B_2$ with weight given by
Eq.\ (\ref{eq:pab1}). This is an exact result in the large-$N$ limit.

For the discussion we 
concentrate on the distribution
$P(\phi')=\int_0^\infty{\rm d}I\,P(I,\phi')$
of the single-mode delay time by itself, which takes the form
\begin{equation}
P(\phi')=\int\limits_0^\infty\int\limits_0^\infty{\rm d}B_1\,
{\rm d}B_2\,
\frac{P(B_1,B_2)(B_2-B_1^2)}{2(B_2+\phi'^2-2B_1\phi')^{3/2}}
\;.
\label{eq:pphi}
\end{equation}

\clearpage

\begin{figure}[ht]
\epsfxsize7cm
\center{\epsfbox{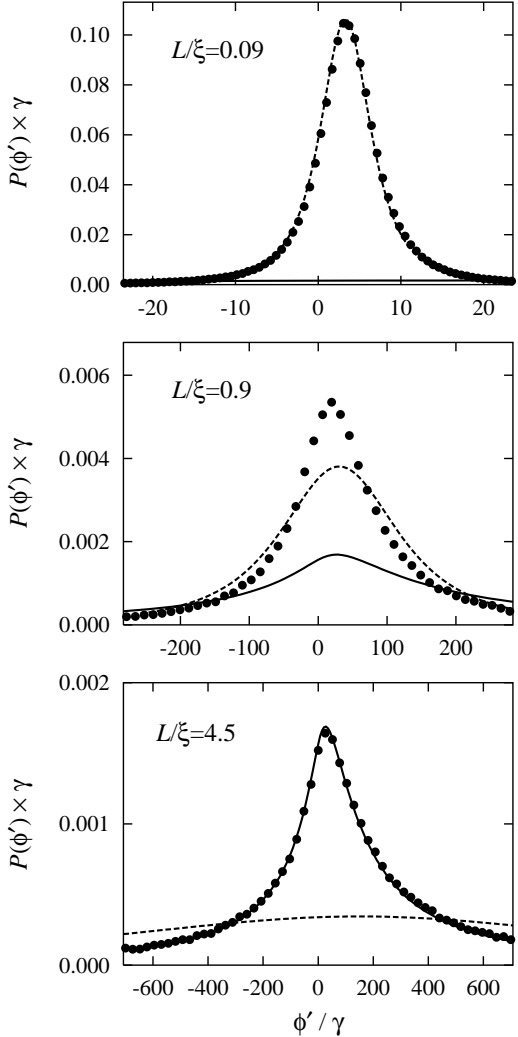}}
\medskip
\caption{Distribution of the single-mode delay time $\phi'$ 
in the diffusive regime (top panel), intermediate regime (middle panel),
and localized regime (lower panel).
The results of numerical simulations (data points)
are compared to the prediction  (\ref{eq:pphidiff})
of diffusion theory
\protect\cite{vantiggelen:1999a,vantiggelen:1999b}
(dashed curve) and the prediction (\ref{eq:pphi})
for the localized regime
(solid curve).
These are results for different incident and detected modes $n\neq m$.
}
\label{fig:num}
\end{figure}

\noindent
We compare this distribution in the localized regime with the
result of diffusion theory
\cite{vantiggelen:1999a,vantiggelen:1999b},
\begin{equation}
P_{\rm diff}(\phi')=
(Q/2\bar{\phi'})[Q+(\phi'/\bar{\phi'}-1)^2]^{-3/2}
\;.
\label{eq:pphidiff}
\end{equation}
In the localized regime the value
$\phi'_{\rm peak}\simeq \gamma N$ at the center of the peak of $P(\phi')$
is much smaller than the width of the peak
$\Delta\phi'\simeq  \gamma N^{3/2}\simeq \phi'_{\rm
peak}(\xi/l)^{1/2}$. This holds also in the diffusive regime, where 
$\phi'_{\rm peak}=\bar{\phi'}$ and
$\Delta\phi'\simeq \phi'_{\rm peak}(L/l)^{1/2}$.
However, the mean $\langle \phi'\rangle=\langle B_1\rangle$ diverges for $P$,
but is finite (equal to $\bar{\phi'}$) for $P_{\rm diff}$.
In the tails $P$ decays $\propto |\phi'|^{-2}$,
while $P_{\rm diff}\propto |\phi'|^{-3}$.

The transition from the diffusive to the localized regime
with increasing $L$
is illustrated in Fig.\ \ref{fig:num}.
The data points are obtained from the numerical simulation 
of scattering
of a scalar wave by a two-dimensional random medium.  The reflection matrices
$r(\omega\pm\frac{1}{2}\delta\omega)$ are computed by applying the method of
recursive Green functions \cite{recgf} to the Helmholtz equation
on a square lattice (lattice constant $a$). The width $W=100\,a$ and the
frequency $\omega=1.4\,c/a$ are chosen such that there are $N=50$
propagating modes. The mean free path
$l=14.0\,a$ is found from the formula
$T=(1+s)^{-1}$ for the transmission probability in the diffusive
regime, where $s=2L/\pi l$ for two-dimensional scattering.
The corresponding localization length $\xi=NL/s=1100\,a$.
The parameter $\gamma=46.3\,a/c$ is found
from $\bar{\phi'}$ in the diffusive regime \cite{note2}. 
The relationship between the parameters $\gamma$, $\bar{\phi'}$, and $Q$
appearing in $P$ and $P_{\rm diff}$ is given by \cite{note1}
\begin{equation}
\label{eq:qp}
\bar{\phi'}=\gamma\frac{s(3+2s)}{3(1+s)}
,\quad
Q=\frac{8s^3+28s^2+30s+15}{5(2s+3)^2}
.
\end{equation}
In Fig.\ \ref{fig:num}, 
the same set of parameters is used for all lengths
to plot the distributions
$P$ (solid curve) and $P_{\rm diff}$ (dashed).
The numerical data agrees very well with
the analytical predictions in their respective regimes of validity.

\begin{figure}
\epsfxsize7cm
\epsfbox{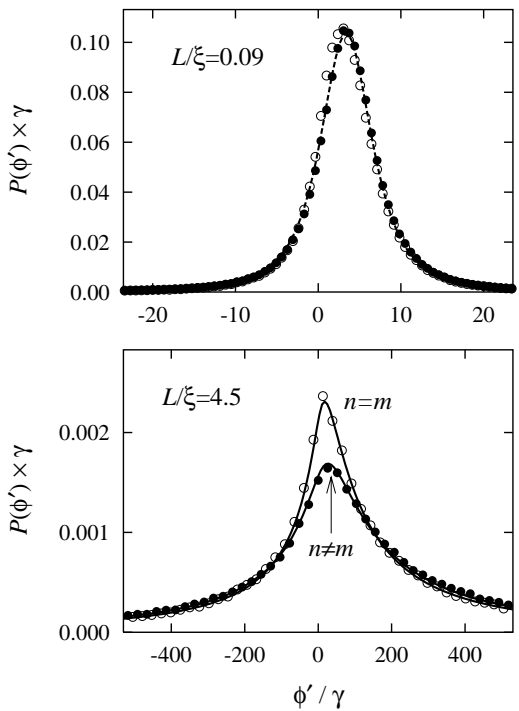}
\medskip
\caption{
Same as Fig. \protect\ref{fig:num}, but now comparing
the case $n\neq m$ of different incident and detected modes (solid
circles) with the equal-mode case $n=m$ (open circles).
The curve for $n=m$ in the lower panel is calculated from
Eqs.\ (\ref{eq:phicd}) and (\ref{eq:pcdnn}).}
\label{fig:pnn}
\end{figure}

We now turn to the  case $n = m$ of equal-mode excitation and detection.
The vectors ${\bf u}$ and ${\bf v}$ in Eq.\
(\ref{eq:corr2}) are then identical, and we can write
\begin{equation}
\label{eq:phicd}
\phi'={\rm Re}\,\frac{C_1}{C_0}
\;,\quad I=|C_0|^2\;,\quad
C_k=\sum_i \tau_i^k u_i^2
\;.
\end{equation}
The joint distribution function
of the complex numbers $C_0$ and $C_1$ can be calculated in the same way
as $P(B_1,B_2)$. We find 
\begin{eqnarray}
P(C_0,C_1)\propto
\exp(-N |C_0|^2/2)
\int_0^\infty{\rm d} x\,
x^2 e^{-x} &&
\nonumber \\
{}\times
\left(1+\frac{|C_1|^2 x^2}{\gamma^2 N^2}-
 \frac{2x}{\gamma N} \,{\rm Re}\, C_0 C_1^*\right)^{-5/2}
\;.&&
\label{eq:pcdnn}
\end{eqnarray}
The maximal value 
$P(\phi'_{\rm peak})=\sqrt{2/\pi N^3\gamma^2}$ for $n=m$
is larger than the maximum of $P(\phi')$ for $n\neq m$
by a factor $\sqrt{2} \times \frac{4096}{1371\pi}=1.35$ in the large-$N$
limit.
This is in contrast to the diffusive regime,
where 
there is no difference in the distributions
of single-mode delay times for $n=m$ and $n\neq m$.
Our analytical expectations
are again in excellent agreement with the numerical simulations,
presented in Fig.\ \ref{fig:pnn}.

In order to explain
the coherent backscattering enhancement of the peak of $P(\phi')$
in qualitative terms,
we compare Eq.\ (\ref{eq:phicd}) for $n=m$ with the
corresponding relation (\ref{eq:corr2}) for $n\neq m$.
The quantities $A_0$ and $A_1$, as well as the quantities
$C_0$ and $C_1$, become mutually independent in the large-$N$ limit.
[The cross-term $(\gamma N)^{-1} \,{\rm Re}\, C_0 C_1^*$ in Eq.\
(\ref{eq:pcdnn}) is of order
$N^{-1/2}$ because $C_0\sim N^{-1/2}$ and $C_1\sim \gamma N$.]
The main contribution to the enhancement of the peak height,
namely the factor of
$\sqrt{2}$,
has the same origin as the 
factor-of-two enhancement of the mean intensity $\bar I$.
More precisely, the relation $P(A_0)=\sqrt{2}\,P(\sqrt{2}\,C_0)$
leads to a rescaling of $P(I)$ for $n=m$ by a factor of $1/2$
[see Eq.\ (\ref{ray})] and to a rescaling of $P(\phi')$ by a factor of
$\sqrt{2}$.
The remaining factor of
$\frac{4096}{1371\pi}= 0.95$ comes from the difference in the
distributions $P(A_1)$ and $P(C_1)$. These distributions turn out to be
very similar, hence the factor is close to unity.
The asymptotic independence of $A_0$ and $A_1$ (as well as of $C_0$ and
$C_1$) is another consequence of the strong fluctuations originating from 
the high density of anomalously large Wigner-Smith
delay times $\tau_i$. In the diffusive regime the corresponding
quantities are strongly correlated, and the 
coherent backscattering enhancement of
the intensity affects both in the same way. Because only their ratio
features in 
$\phi'$, this effect cancels and no difference is observed in
$P_{\rm diff}(\phi')$ for $n=m$ and $n\neq m$.

In conclusion, we have discovered a dynamical effect of coherent
backscattering, that requires localization for its existence.
Computer simulations confirm our prediction, which now awaits
experimental observation.

We thank P. W. Brouwer for valuable advice. This work was 
supported by the Dutch Science Foundation NWO/FOM.


\begin{references}
\bibitem{note2}
This value of $\gamma$ is somewhat bigger than
the value $\gamma= \pi^2 l/4c=34.5\,
a/c$ for two-dimensional scattering, as a consequence of the anisotropic
dispersion relation on a square lattice.
\end{references}
\end{document}